\documentclass[10pt,preprint]{emulateapj}

\newcommand{\mymail}{oysteol@astro.uio.no}
\begin{document}

\altaffiltext{1}{also at Center of Mathematics for Applications, University of 
Oslo, P.O. Box 1053 Blindern, N-0316 Oslo, Norway}

\title{Velocities measured in small scale solar magnetic elements.}

\author{{\O}ystein Langangen}
\affil{Institute of Theoretical Astrophysics, University of Oslo, 
P.O. Box 1029 Blindern, N-0315 Oslo, Norway}
\email{\mymail}
\author{Mats Carlsson \altaffilmark{1}}
\affil{Institute of Theoretical Astrophysics, University of Oslo, 
P.O. Box 1029 Blindern, N-0315 Oslo, Norway}
\author{Luc Rouppe van der Voort \altaffilmark{1}}
\affil{Institute of Theoretical Astrophysics, University of Oslo, 
P.O. Box 1029 Blindern, N-0315 Oslo, Norway}
\and
\author{R. F. Stein}
\affil{Department of Physics and Astronomy, Michigan State University, 
East Lansing, MI 48823}
\begin{abstract}
We have obtained high resolution spectrograms of small scale magnetic
structures with the Swedish 1-m Solar Telescope. We present Doppler
measurements at $0\farcs{2}$ spatial resolution of bright points,
ribbons and flowers and their immediate surroundings, in the 
C~{\small{I}} 5380.3~{\AA} line (formed in the deep 
photosphere) and the two 
Fe~{\small{I}} lines at 5379.6~{\AA} and 5386.3~{\AA}.
The velocity inside the flowers and ribbons are measured
to be almost zero, while we observe downflows at the edges.
These downflows are increasing with decreasing height.
We also analyze realistic magneto-convective simulations to obtain a
better understanding of the interpretation of the observed signal. We calculate 
how the Doppler signal depends on the velocity field in various structures.
Both the smearing
effect of the non-negligible width of this velocity response function along
the line of sight and of the smearing from the telescope and atmospheric
point spread function are discussed. These studies
lead us to the conclusion that the velocity inside the magnetic elements are 
really upflow of the order 1--2~km\,s${}^{-1}$ while the downflows at 
the edges really are much stronger than observed, 
of the order 1.5--3.3~km\,s${}^{-1}$.
\end{abstract}


\section{Introduction}
\label{intro}

In the Sun's photosphere one can observe magnetic structures in a range of 
different scales. From sunspots of tens of Mm size to small scale 
magnetic elements of about 200~km or less. 
Observationally these small scale magnetic elements
become visible as features brighter than the
surroundings (leading to the often used term ``bright points'') when the 
photosphere is imaged at sufficiently high angular resolutions \citep{1973Dunn}.
\citet{TitleRes} showed that bright points
can not be detected with spatial resolutions worse than about $0\farcs4$ due
to a smearing out of the contrast because bright points are surrounded by
darker areas.
As observational methods have become more sophisticated, such as adaptive 
optics \citep[e.g.][]{2000Rimmele,SSTAO} 
and post-processing methods \citep[e.g.][]{Speckle,momfbd},
the observations of such structures have become almost routine.
Recently \citet{BergerMag} concluded that at 100 km resolution the magnetic 
elements 
do not always resolve into discrete bright elements. Instead, in plage
regions of stronger average magnetic field, the small scale 
magnetic elements are found to be concentrated into elongated ``ribbons'' and
round ``flowers'' with a darker core and bright edges. These elements are 
constantly evolving; the merging and splitting of ``flux sheets'' and the 
transitions between the ribbons, flowers, and micro-pores are described by 
\citet{RouppeMag}.
The understanding of these small scale elements is  
of great importance and these structures have therefore been subjected to 
intense research in recent years \citep[e.g.,][]{2001Kiselman,
2001Rutten,2001Almeida,2001Steiner,2004Keller,Mats3d}. 
This has resulted in good
theoretical models and a good understanding of why the magnetic elements
look the way they do in observations.
One main conclusion of the above papers is that the dominant reason for the 
brightness of bright 
points is the lower density inside the bright point due to magnetic 
pressure, hence allowing the observer to look into deeper layers. These layers
are hotter than the higher layers seen outside the magnetic elements because
of the temperature increasing with depth. At equal geometric height, the
temperature inside the magnetic element is lower than in the surroundings
due to the suppression of convective energy transport by the magnetic field.
This temperature contrast may be large enough in larger magnetic concentrations
that it offsets the effect of the difference in formation height and the
magnetic element no longer looks bright. Smaller magnetic elements 
(bright points)
and the edges of the larger magnetic elements (ribbons, flowers, pores) are
heated radiatively by the surrounding hotter, non-magnetic, atmosphere and
we get a bright point or a bright edge.
Bright points have often been observed in the G-band (spectral domain 
dominated by CH-molecular lines at 4300~{\AA}). This is because bright 
points have higher contrast in this domain due to an increased 
height difference of the optical depth unity height because of the destruction
of CH-molecules in low-density magnetic elements. 

\citet{2004Rimmele} presents measurements of the line of sight (LOS) velocity 
field in and around magnetic flux concentrations
based on narrow-band filtergrams. He finds downdrafts at the 
edge of flux concentrations of ``a few hundred meters per second.''
The size of these downflow areas is approximately ${0\farcs 2}$,
which indicates that they are smaller since this is the approximate
spatial resolution of the observations. Furthermore he concludes that the 
downdraft 
gets narrower and stronger in deeper layers and the plasma within the 
flux concentration is more or less at rest. 
\citet{RouppeMag} found that magnetic structures such as flowers, 
ribbons and flux sheets have a weak upflow inside and a sharp downdraft 
at the edge in accordance with \citet{2004Rimmele}. 
The LOS flow inside flowers and ribbons was measured to be about 
0--150~m\,s${}^{-1}$ upflow and the downdrafts at the edge to about 
360~m\,s${}^{-1}$. 

Both \citet{2004Rimmele} and \citet{RouppeMag} use filtergrams at only two 
spectral positions
to derive the velocities --- adding more spectral sampling points 
would increase the accuracy of the measurements and would also make
it possible to derive the velocity as a function of height. Spectroscopic 
measurements of bright points were reported by \citet{2002Langhans} and they
derive velocity maps showing downflows close to bright points.  
Furthermore \citet{2004Langhans} use spectroscopic diagnostics to 
measure velocities inside bright points, at resolutions of about
0\farcs{3} they find both upflows and downflows inside bright points.

In this paper, we present spectroscopic data of magnetic elements such as  
ribbons, flowers, and bright points with excellent spatial resolution. 
Intensity cuts of the smallest features and analysis of the spatial power
spectrum show that the spatial resolution is better than 0\farcs{2} for
the spectra used in this work.
These spectra are used to obtain the LOS velocities in and close 
to small scale magnetic elements. Furthermore we analyze a three dimensional 
simulation
of magneto-convection by solving the radiative transfer through the
simulation cube in the same spectral domain as in the observations.
These simulations are used to understand the effects of the smearing
along the line of sight (the velocity response function) and spatial smearing
(the instrument and atmospheric point spread function(PSF)) have on
the determinations of velocities.

In \S~\ref{Obs} we present and discuss the observational program and the 
instrumentation. The data reduction methods are discussed in \S~\ref{Data}. 
In \S~\ref{Doppler} we present the results of the Doppler 
shift measurements in the three spectral lines. The analysis of the
Magneto-convective simulation, the determination of the velocity response
functions in the diagnostics used, and the discussion of the smearing effects 
this
kind of velocity measurements are hampered with  are presented in 
\S~\ref{Sim}. 
Finally, we summarize our results in \S~\ref{Con}.

\section{Observing program and instrumentation}
\label{Obs}
\begin{figure*}[!ht]
\includegraphics[width=\textwidth]{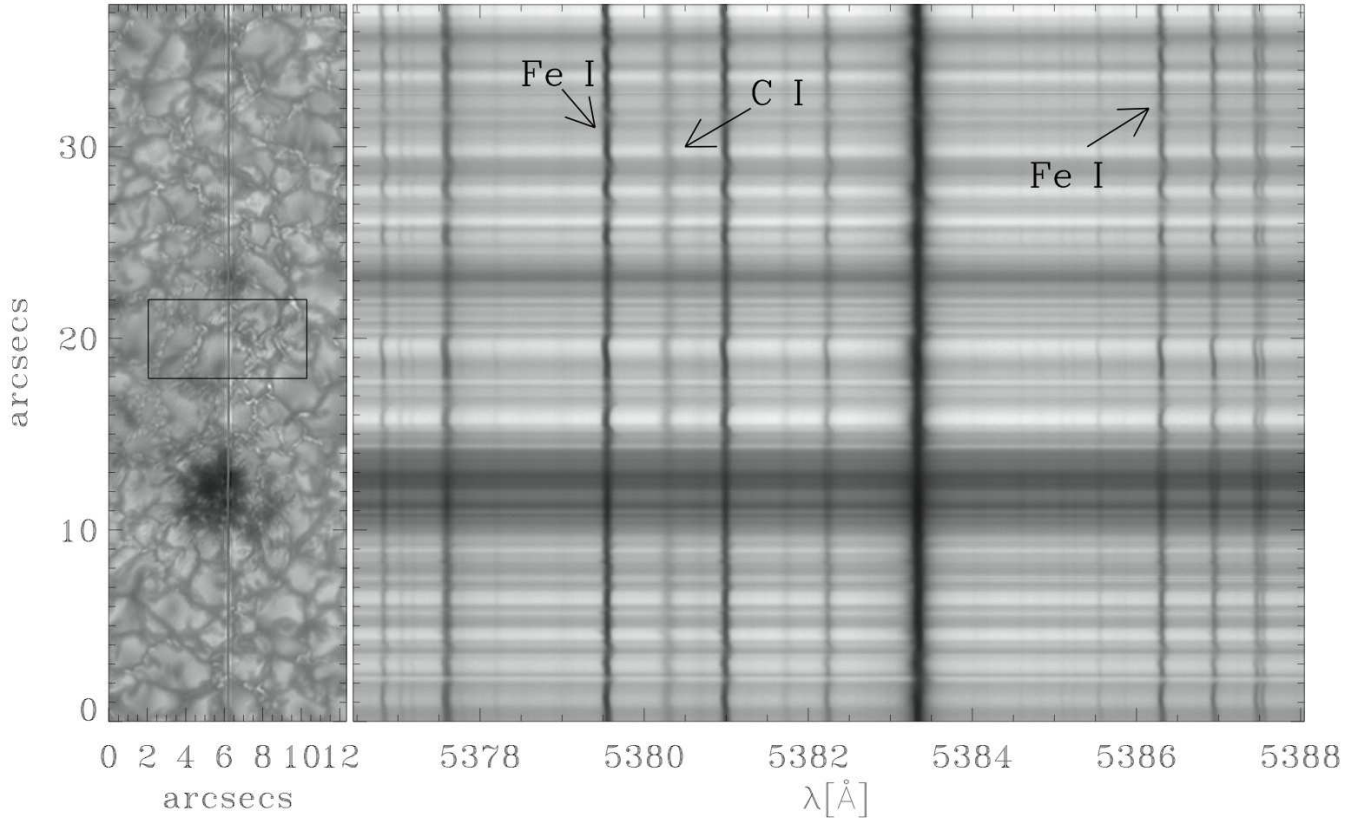}
\caption{Slit-jaw image ({\it left}) and  
the corresponding
spectrogram ({\it right}). The slit-jaw image has been MFBD processed
to facilitate the identification of small scale structures. 
The box in the slit-jaw image 
marks the area which has been magnified in Fig.\ref{plotthree}} 
\label{plotone}
\end{figure*}

The main aim of these observations was to obtain good spectra in and close
to small scale magnetic structures. On 2005 May 13 we observed
the NOAA 
active region 10759 ($N8.4^\circ$, $E10.5^\circ$, $\mu=0.97$)
using a small pore as adaptive optics lock point, see 
Fig.\ref{plotone} for an example of a slit-jaw image.
 
The spectra were obtained using the Swedish 1-m Solar Telescope (SST) 
\citep{SST}
together with the adaptive optics system \citep{SSTAO} and the TRI-Port
Polarimetric Echelle-Littrow (TRIPPEL) spectrograph. As indicated by the name,
the TRIPPEL spectrograph is a Littrow spectrograph with an Echelle 
grating using 79 grooves mm${}^{-1}$. The blaze angle is $63.43^\circ$, 
the focal length is $1500$ mm, and the slit width is $25~{\micron}$. 
Simultaneous spectra of two wavelength intervals, 4566--4576~{\AA} (hereafter
called
the 4571 interval) and 5376.5--5388~{\AA} (the 5380 interval), 
were obtained. In the present paper we focus on the 5380 interval since these 
spectra were obtained with the shortest exposure time and thus the Doppler 
shifts in the spectral lines will be less affected by seeing.
The spectrograph should always be operated with angles close to
the blaze angle. We observe at $62.73^\circ$ and in order 47 and 42 for the 
4571 and 5380 intervals respectively. The theoretical slit-limited 
bandpass for the 5380 interval is $W_\lambda=23$ m{\AA} or in terms of 
effective theoretical spectral resolution $R\approx 234000$.
In all our comparisons with FTS spectra and simulations we convolve 
these spectra to this resolution.

Two Megaplus 1.6 cameras (KAF-0401E Blue Plus and KAF-0401 Image sensor)
were used at exposure times of 200 ms 
and 80 ms for the 4571 interval and 5380 interval respectively. The two cameras 
have a quantum efficiency of about $35\%$ and $25\%$ at the operating 
wavelengths. 
In addition to the spectral cameras we used two cameras to obtain slit-jaw 
images. One of the slit-jaw cameras using a 10.3~{\AA} bandpass interference
filter centered at 4572.6~{\AA} was slaved to the 
spectral camera at the same wavelength.
The other slit-jaw camera, using a 9.2~{\AA} bandpass interference filter 
centered at 5321~{\AA}, had exposure times of 8 ms and was not slaved.
In the spatial domain the sampling is 0\farcs{0411} pixel${}^{-1}$.
With this setup we observed several series of good to excellent seeing
with a fixed slit-position. 
One of these series, with a duration of about 17 minutes 
(10:59:40--11:17:06 UT),
is excellent both with respect to seeing conditions and structures crossed 
by the slit, see Fig.\ref{plotone}.
In the following all the data presented come from this series and we 
consider data of excellent quality spread over the full 17 minutes.

\section{Data reduction}
\label{Data}
\subsection{Flat fields and dark currents}

The spectrograms are corrected for dark currents and flat fields.
Flat fields are constructed from 5 series of 50 images of random scans of 
the quiet solar disk center. 
We then construct a mean spectrum. To obtain a mean spectrum we need to 
correct for the spectrograph's main distortions, namely smile (the curvature 
of the spectral lines) and keystone (different spectral dispersion at 
different slit positions). 
The distortions are determined by fitting a 4th-order polynomial 
to the central part of the spectral line and in this way we measure the
line core position over the slit. This is done for five lines distributed 
over the spectrogram. 
The smile is of order 4 pixels while the keystone is of order 0.15 pixels, thus
the smile is the most profound aberration. 
The final flat fields are obtained by dividing the total of 
the 250 images by a distortion corrected mean spectrogram. 
In this way we remove both
spatial and spectral information without losing the fixed CCD dependent 
signal and without any interpolation of pixels in the flat field.

\subsection{Wavelength calibration}

\begin{figure}[!!ht]
\includegraphics[width=0.5\textwidth]{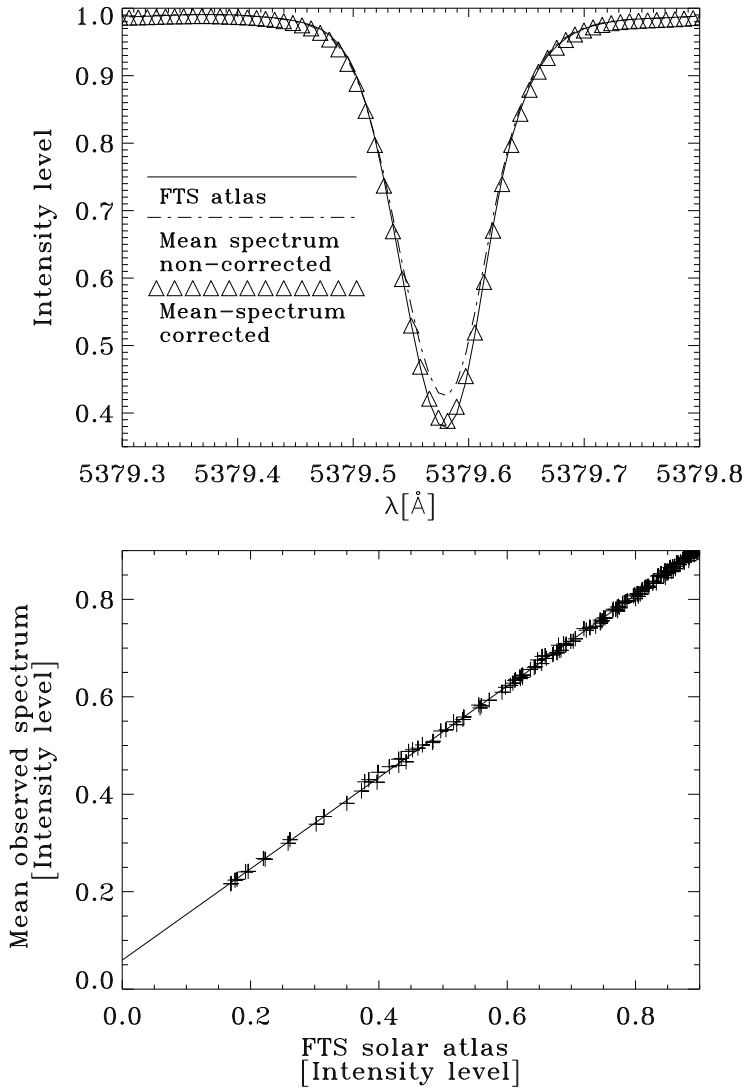}
\caption{{\it Upper:} FTS atlas of the Fe~\small{I}
\normalsize 5379.6~{\AA} line ({\it solid}) and the mean spectrum 
({\it dashed-dotted}). The scattered light corrected mean spectrum is showed 
with triangles.
{\it Lower:} Intensity level (intensity relative to the mean continuum
intensity) of our quiet Sun spectrum (mean flat field spectrum) versus
the intensity level of the FTS atlas ({\it crosses}) 
and a least-square linear fit to these data ({\it solid}). 
}
\label{plottwo}
\end{figure}

We make a mean solar spectrogram by adding together all 250 flat field spectra. 
The mean aberration corrected spectrum is then compared to the FTS atlas of 
\citet{1987Brault}. This atlas has proved to be well calibrated in wavelength
and it shows no systematic offset in line shifts with wavelength 
\citep{1998Allende}. Since the solar atlas is corrected for the Earth's 
rotation, the Earth's orbital motion and the Sun's rotation we automatically 
get a corrected spectrogram when we calibrate using the FTS atlas. 
The upper panel of Fig.\ref{plottwo} gives a good impression of the 
accuracy of this calibration. Note that in this way our wavelength scale 
is the same 
as that of the FTS atlas; there is thus no correction for gravitational 
redshift. The slit covers a region of almost 38\arcsec{} or approximately 28 
Mm on the Sun. 
During the observations there is a temporal change of the spectral position
on the cameras. This wavelength shift is the same in both spectral regions.
Part of the shift is periodic, with an amplitude of 50--150 m\,s${}^{-1}$ and 
a period of five minutes. This variation is caused by global oscillations 
averaged over the slit length. In addition there are slow trends, probably 
caused by temperature variations in the spectrograph. The frame to frame 
standard deviation is 8~m\,s${}^{-1}$.
We correct for this temporal change by applying a shift to the wavelength 
calibration determined form the flat field spectra.
Since the slit is 
covering an active region, the mean spectrum of each slit is not directly 
comparable to the solar atlas \citep[][and references therein]{1990Brandt}.
\citet{1990Brandt} measure the mean convective blueshift in 19 Fe~{\small{I}} 
\normalsize lines, including the Fe {\small{I}}~5379~{\AA} line, 
in quiet Sun and magnetically active regions. They find that the convective 
blueshifts are decreased from 350~m\,s${}^{-1}$ in quiet Sun to 
120~m\,s${}^{-1}$ in active areas. 
We determine the temporal shifts by comparing the line center of the
mean Fe~{\small{I}}~5379~{\AA} line in each spectrogram (averaged over an active
region) with the line center position of the quiet Sun atlas. We thus
need to compensate for the difference in line center convective
blueshift between active regions and the quiet Sun. The difference
of 230~m\,s${}^{-1}$ found by \citet{1990Brandt} refers to an intensity level
of 60\% and not line center. From our observations we find that the difference
is 30~m\,s${}^{-1}$ smaller at line center. We thus apply a redshift of 
200~m\,s${}^{-1}$
to the atlas before comparing with our mean Fe~{\small{I}}~5379~\AA~line when
determining the temporal shift.
 
Instead of correcting each spectrum for smile and keystone we use the
determined aberrations to determine a separate wavelength scale for each 
slit position. This procedure avoids interpolation of the spectrogram data.

\subsection{Scattered light}

When the mean quiet Sun spectrum is plotted against the FTS atlas, 
see Fig.\ref{plottwo} upper panel, it is clear that the mean spectrum has 
shallower lines than the FTS atlas. This discrepancy is due to scattered light,
mainly from the diffuse scattering in the spectrograph. In the lower panel in 
Fig.\ref{plottwo} we have plotted 
the intensity level of the mean spectrum against the FTS intensity level.
We use a least-square linear fit to these data points.
Assuming a constant level of diffuse scattered light, the offset and the new 
continuum calibration of the mean spectrum is given by this fit. 
The amount of constant scattered light is 6~\%. This level is subtracted from 
the spectra and the resulting spectra are then divided by a scaling factor 
found from the slope of the curve in the lower panel of Fig.\ref{plottwo}. 
Again we refer to Fig.\ref{plottwo} upper panel to see the effect of this 
correction of the scattered light. 

According to 
\citet{1994Sobotka}, ``the mean intensity in abnormal granulation is equal to
that of quiet granulation,'' so we expect the mean spectrum from the flat fields
to have the same mean count as we have in each 
of the mean spectrograms when we do not include the pores. This turns out not
to be the case. There are many reasons for this discrepancy, different 
atmospheric conditions, the changing path length of the light ray 
in the atmosphere due to the time of the day and changing intensity on the 
Sun due to solar oscillations (p-modes).  
All these effects are corrected for by introducing a
scaling factor which gives the same mean solar continuum intensity in the
mean spectrum of each spectrogram as in the mean solar atlas. We 
multiply with this scaling factor before we correct for scattered light.

\subsection{Slit-jaw images}

The slit-jaw images are corrected for dark currents and flat fields
and post-processed using the multi frame
blind deconvolution (MFBD) method \citep{momfbd}. The intensity on the slit in
the slit-jaw image is correlated to the mean intensity in the 
corresponding spectrogram to obtain the relative scaling and offset between
the spectrogram and the slit-jaw image. The aligned slit-jaw images are used
to identify the structures crossed by the slit, see Fig.\ref{plotone}.
Note that the slit-jaw image has shorter exposure time than the spectrum
and has been post-processed --- it is thus sharper and has a more narrow
point-spread-function than the spectrogram.

\section{Observed Doppler shifts}
\label{Doppler}
The 5380 interval contains the deep forming \citep{1977Livingston} 
C~{\small{I}} 5380.3~{\AA} line, see Fig.\ref{plotone}. 
Furthermore it contains a well suited quite strong Fe~{\small{I}} line 
at 5379.6~{\AA} and a weaker Fe~{\small{I}}
line at 5386.3~{\AA}. These two iron lines have been singled out by 
other authors as good both in the sense that they are blend free and thus 
are well suited to measure the line asymmetry \citep{1981Dravins}
and that their central wavelengths are well known \citep{1998Allende}. 
\citet{1994Nave} give the central wavelength of these Fe~{\small{I}} 
 lines, see Table~\ref{wavelength}, and they claim the error in the central 
wavelength of these two lines to be less than 1.25~m{\AA} and 2.5~m{\AA} 
respectively, which corresponds to uncertainties in velocities of 
70~m\,s${}^{-1}$ and 140~m\,s${}^{-1}$. 
The central wavelength of the C~{\small{I}} line is given by 
\citet{1966Johansson} and he quotes the uncertainty to be less than
0.02~{\AA}, which corresponds to 1115~m\,s${}^{-1}$. 
This high uncertainty makes it difficult to use this line to determine
velocities on an absolute scale.
The Fe~{\small{I}} line at 5383.3~{\AA} was also picked out by
\citet{1981Dravins}, but we do not use this line due to a blend in the 
blue wing.
We use the laboratory wavelengths as reference wavelengths for the two
Fe {\small{I}} lines, while we use a reference wavelength which
gives us a more realistic convective blueshift for the C~{\small{I}} line. 
Using the wavelength given by \citet{1966Johansson}, 5380.337~{\AA}, 
we get a convective blueshift of 1465~m\,s${}^{-1}$. 
Instead we use a reference wavelength which is in agreement with a 
3D magneto-convective simulation, see \S~\ref{Sim}.
The simulation gives a convective blueshift of 664~m\,s${}^{-1}$, which
indicate a convective blueshift in non-magnetic regions of 864~m\,s${}^{-1}$,
which corresponds to a reference wavelength of 5380.3262~{\AA}.
This wavelength is well within the given uncertainty of 0.02 {\AA}.
The line-core wavelengths in the FTS atlas, the reference wavelengths, 
used to calculate the LOS velocities, and the corresponding 
blueshifts are shown in Table~\ref{wavelength}. Notice that we have corrected
for a gravitational redshift of 633 m\,s${}^{-1}$.
The Land{\'e} g-factors of the Fe~lines are $1.10$ and $1.17$, for the 
C-line it is unknown but greater than zero.
 
\begin{table}[!ht]
\begin{tabular}{rrrr}
\multicolumn{4}{c}{Central wavelengths and convective blueshifts}\\
\tableline
\tableline
Atomic& Atlas& Reference& Velocity\\
species&[\AA]&[\AA]&[m\,s${}^{-1}$]\\
\tableline
Fe~{\small{I}} & 5379.5800&5379.5740&-307\\ 
C~{\small{I}} & 5380.3221 &5380.3262&-864\\ 
Fe~{\small{I}} & 5386.3349&5386.3341&-589\\
\tableline
\end{tabular}
\caption{}
\label{wavelength}
\end{table} 
 
Since the C~{\small{I}} line is very shallow we only determine 
the total line shift for this line, while the Fe~{\small{I}} 
lines are well suited for determining the 
Doppler shift at different intensity levels by determining the 
bisector of the line. To facilitate a comparison of velocities in different
features we use the same absolute intensity levels in the various features
(we thus do {\em not} use the local continuum of that feature). The
intensity levels used are relative to the average continuum level of
the average Sun at disk center. This continuum value, as given by 
\citet{1984Neckel}, and the values at other intensity levels are given in 
Table \ref{inttemp}, both as absolute intensities and as radiation temperatures.
The latter is indicative of the gas temperature at the level of formation
of the radiation at this intensity level.
 
\begin{table}[!ht]
\begin{tabular}{rrr}
\multicolumn{3}{c}{Intensity level, specific intensity,}\\
\multicolumn{3}{c}{and radiation temperature.}\\
\tableline
\tableline
IL& I$_\nu$&RT\\
 &[erg\,cm${}^{-2}$str${}^{-1}$\,s${}^{-1}$\,Hz${}^{-1}$]&[K]\\
\tableline
  1.00 &   3.70E-05 &     6293 \\
  0.95 &   3.51E-05 &     6219 \\
  0.90 &   3.33E-05 &     6143 \\
  0.85 &   3.14E-05 &     6064 \\
  0.80 &   2.96E-05 &     5983 \\
  0.75 &   2.77E-05 &     5899 \\
  0.70 &   2.59E-05 &     5811 \\
  0.65 &   2.40E-05 &     5720 \\
  0.60 &   2.22E-05 &     5625 \\
  0.55 &   2.03E-05 &     5524 \\
  0.50 &   1.85E-05 &     5418 \\
\tableline
\end{tabular}
\caption{}
\label{inttemp}
\end{table}
 
The total line shift in the C~{\small{I}} line
is calculated using a Gaussian fit to the line.
When we calculate the velocities we correct for gravitational redshift.

\begin{figure}[!ht]
\includegraphics[width=0.5\textwidth]{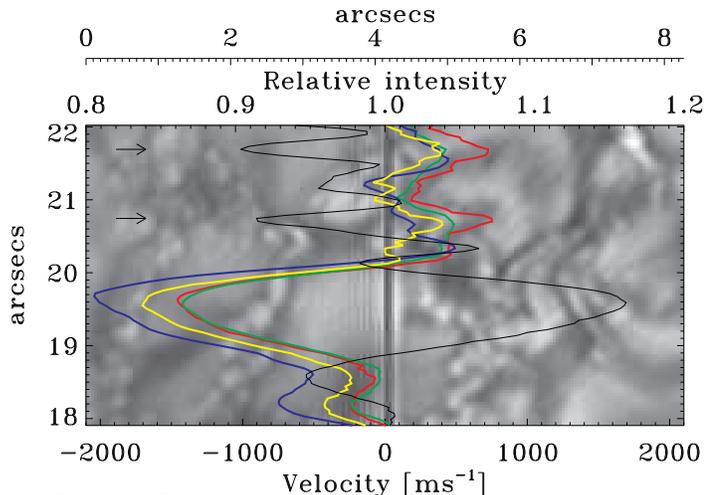}
\caption{One of the snapshots shows the spectrograph's slit crossing a 
flower/ribbon 
(20\farcs{8}--22\farcs{7}) and a granule (18\farcs{5}--20\farcs{0}). 
The velocities are shown as measured in the Fe~{\small{I}} 5379.6~{\AA} line
at intensity level 0.85 ({\it red}) and 0.65 ({\it green}),
in the
C~{\small{I}} line ({\it blue}) and 
in the Fe~{\small{I}} 5386.3~{\AA} line
at intensity level 0.85 ({\it yellow}).
Positive velocities mean downflow. The black line shows the relative continuum 
intensity in the spectrogram. The two IGMs are marked with arrows.}
\label{plotthree}
\end{figure}

Our slit crosses from two to three ribbons, depending
both on the temporal evolution of the photosphere as well as the changes in slit
position due to differential seeing. It must be emphasized that we do not have
magnetic information so all classification has to be done 
purely by visual inspection. The kind of ribbon-structures we see 
has been shown to be magnetic \citep{BergerMag}. A flower/ribbon and a typical 
granule can be seen in Fig.\ref{plotthree}. This Figure 
also shows velocities and the continuum intensity at the slit
position. It is clear that we have downflows at the edge of the flower/ribbon 
(the two arrows), which are increasing with increasing intensity level 
in the line (formation deeper in the atmosphere).
Furthermore, the velocity in the granule is somewhat high 
(the mean measured value in our observations is about 1.2--1.7~km\,s${}^{-1}$), 
with about 1.4--2.0~km\,s${}^{-1}$ upflow. 
Finally, we see that in the center of the flower/ribbon
we measure the velocities to be more or less zero. 
Intensity minima between magnetic structures and granules are found
by visual inspection, these are hereafter called 
``intergranular lanes close to magnetic structures'' (IGM).
The peak velocity close to the 
intensity minimum in the IGM is measured, this peak can vary slightly from the 
intensity minimum, but not more than $0\farcs{15}$, see Fig.\ref{plotthree}.
This peak velocity is measured for up to 51 different IGMs, the velocities 
and the standard deviations are shown in Table~\ref{vel}. 
The pixel--to--pixel standard deviation in the velocity determination is 
40~m\,s${}^{-1}$.

\begin{table}[!ht]
\begin{tabular}{rrrrr}
\multicolumn{5}{c}{Velocities in IGMs}\\
\tableline
\tableline
Atomic&IL& Velocity&$\sigma$ & Samples\\
species& & [m\,s${}^{-1}$] & [m\,s${}^{-1}$] & \\
\tableline
C~{\small{I}} 5380 &$\ldots$& 191 & 227& 51\\
Fe~{\small{I}} 5379 &0.90&755&326&51\\
&0.85&563&225&51\\
&0.80&459&188&51\\
&0.75&407&179&51\\
&0.70&364&173&51\\
&0.65&334&183&51\\
&0.60&314&182&47\\
&0.55&283&179&35\\
Fe~{\small{I}} 5386&0.90&410&273&51\\
&0.85&278&193&51\\
&0.80&269&216&44\\
\tableline
\end{tabular}
\caption{}
\label{vel}
\end{table}

A statistical analysis of the minimum LOS velocity within flowers and 
ribbons (Ribbon centers or RC) gives us velocities around zero~m\,s${}^{-1}$ 
in the Fe~{\small{I}} 5379 {\AA} line, while the 
the C~{\small{I}} line shows an up flow of about 300~m\,s${}^{-1}$.
The Fe~{\small{I}} 5386 {\AA} shows a small upflow 
inside RCs with a mean of about 200~m\,s${}^{-1}$.
The standard deviations are of the order 100--200~m\,s${}^{-1}$ in the
Fe~{\small{I}} lines, while the C~{\small{I}} line
has a standard deviation of about 350~m\,s${}^{-1}$.
These results are hampered with relatively high uncertainties of a 
few hundred~m\,s${}^{-1}$.
We therefore want to investigate the effects of smearing along the 
line of sight and spatially on these velocity measurements. 

We have also measured velocities inside bright points but since the spatial
extent of each bright point is small they are difficult to identify.
Nevertheless we visually classify 15 bright points and measure the velocities 
at intensity maximum. We measure a mean downflow of a few hundred m\,s${}^{-1}$,
but the velocities inside bright points vary a lot, and we also measure
upflows inside some of the bright points.
The standard deviation is quite high, about 0.5~km\,s${}^{-1}$. Due to the 
small spatial extent of each bright point, the measured velocities are strongly 
affected by the surrounding intergranular velocity.

The effect of a non-zero g-factor is increased broadening in magnetic
areas. For a non-symmetric profile this broadening may also introduce
a change in the determined bisector.  We have quantified this effect
by applying a Zeeman splitting corresponding to a magnetic field
strength of 1000 Gauss to the observed profiles and redetermined the
bisector. The change in measured velocity is less than 40 m\,s${}^{-1}$.

\section{Simulations}\label{Sim}

Since the intensity level can be considered as a height parameter it
is in principle possible to obtain the velocity as a function of absolute 
physical height. In order to explore this possibility we need a good model 
with which we can calibrate the velocities with height. 
One very suitable model is the magneto-convective simulation used by 
\citet{Mats3d}. 
In this realistic simulation we can see bright points as well as more 
elongated magnetic structures. To tailor this simulation to our observational 
setup we use 12 atmospheric snapshots, including the same snapshot as the one 
used by \citet{Mats3d}, with one minute time difference between the
snapshots. We solve the equations of radiative transfer in LTE for these
3D atmospheres, using the MULTI code \citep{multi}, for
1364 frequency points 
spanning from 5377.6~{\AA} to 5389.4~{\AA}. A line list of 249 lines is used. 
The line data was retrieved from the Vienna Atomic Line Database (VALD) 
\citep{vald,1995Piskunov,1999Ryabchikova}, with a few exceptions;
the C~{\small{I}} oscillator strength is taken from 
\citet{1993Hibbert} and an Fe line at 5382.5~{\AA} was removed due 
to obvious errors in its central wavelength and oscillator strength. 
 
\begin{figure}[!ht]
\includegraphics[width=0.5\textwidth]{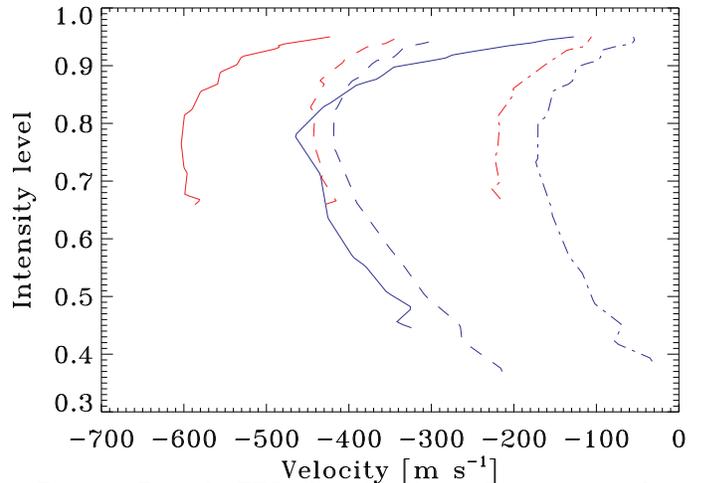}
\caption{From the FTS solar atlas, the total magnetic simulation and the total 
non-magnetic simulation we derive the bisectors in the two Fe~{\small{I}}
lines. The mean bisectors in the Fe~{\small{I}} 5379 line ({\it blue}) and 
the Fe~{\small{I}} 5386~{\AA} line ({\it red}) in the 
FTS atlas ({\it solid}), the non-magnetic simulation ({\it dashed}), 
and in the magnetic simulation ({\it dashed-dotted}) are shown. 
The shift in the Fe~{\small{I}} 5386~{\AA}~line indicate 
that there might be an error in the laboratory wavelength of 
about 100~m\,s${}^{-1}$.  Furthermore the laboratory wavelength of the
Fe~{\small{I}}~5379~{\AA} seems to be well determined.
Also notice the reduced blueshifts in the magnetic simulation of approximately 
200~m\,s${}^{-1}$, which are in good agreement with observational results.}
\label{plotfour}
\end{figure}
 
A temporally and spatially averaged mean spectrum over 12 minutes and 
$8\farcs{27} \times 8\farcs{27}$, totally over 750000 spectra, is constructed. 
Since the simulation is covering an active region we expect
the mean spectrum to have less blueshift, have shallower lines 
and to have more vertical lower bisectors (intensity level less than 0.7)
than similar mean spectra obtained in non-magnetic regions 
\citep[][and references therein]{1990Brandt}. 
For reference we therefore solve the equations of radiative transfer in LTE in 
three non-magnetic snapshots with 1 minute cadence, 
totally over 190000 spectra. Even though the temporal 
sampling is quite low we believe that the difference from a bigger temporal 
sample will be within 1~m{\AA} or approximately 50~m\,s${}^{-1}$.
In the non-magnetic simulation we have deeper strong Fe~lines than the atlas, 
this is due to NLTE effects \citep{2001Shchukina}.
The Fe~lines in the mean magnetic spectrum do not show any reduced line depth.
This might be because the magnetic field in
the simulation is too weak but also because of NLTE effects. 
The convective blueshifts are in general well reproduced in the non-magnetic 
simulation \citep{2000Asplund1}. One difference is that the upper
part of the bisector shows larger blueshift than the atlas, see Fig.\ref{plotfour}.
We therefore use the blueshift in the magnetic simulation as reference 
wavelength for the C~{\small{I}} line since this simulations seems 
to give a more realistic convective blueshift in this line, 
see \S~\ref{Doppler}.
The magnetic simulation shows a blueshift that is about 
200 m\,s${}^{-1}$ smaller
than the non-magnetic simulation, very similar to what was found observationally
by \citep{1990Brandt}.
Furthermore the lower part of the bisector in the magnetic simulation 
is more vertical than in the non-magnetic simulation, also similar to
observations.
These are indications that the simulations are reproducing the velocities
in a realistic manner.

\begin{figure}[!ht]
\includegraphics[width=0.5\textwidth]{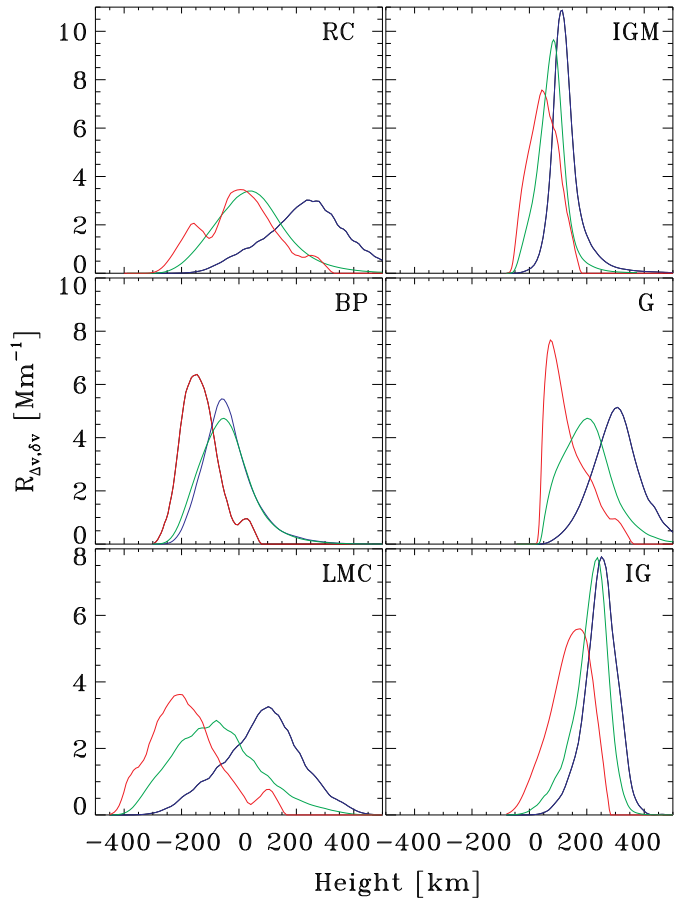}
\caption{Some typical response functions to velocity.
The response function of the total C~{\small{I}} 
line Doppler shift({\it red}), and of the line core Doppler shift for the 
Fe~{\small{I}}~5379.6~{\AA} line ({\it blue}) and Fe~\small{I} 
\normalsize~5386.3~{\AA} line ({\it green}) are shown. 
The different panels show from upper left to lower right: Ribbon 
center (RC), Intergranular lane close to Magnetic element (IGM), 
Bright point (BP), Granule (G), Large Magnetic Concentration (LMC), 
Inter granular lane (IG).
The different points are defined in Fig.\ref{plotseven}, the 
BP, G, LMC, and IG are the same points as those used by \citet{Mats3d}.}
\label{plotfive}
\end{figure}

\begin{figure}[!ht]
\includegraphics[width=0.5\textwidth]{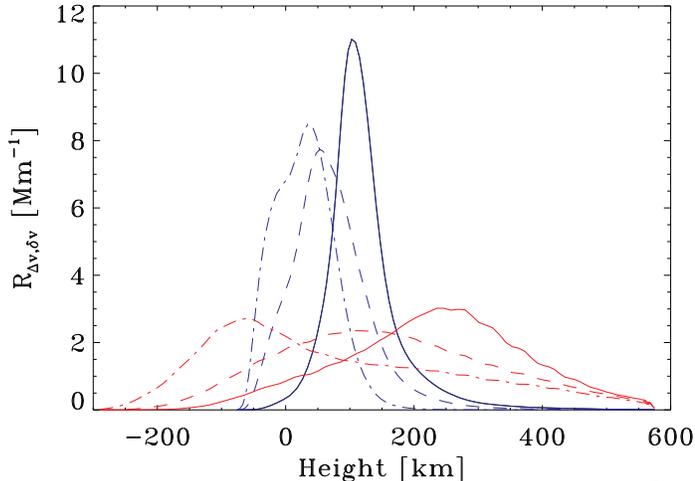}
\caption{Height variation of the response functions in the Fe~{\small{I}}
5379~{\AA} line in 
the Ribbon center (RC) ({\it red}) 
and the Intergranular lane close to Magnetic
element (IGM) ({\it blue}). 
The intensity levels are 0.4 ({\it solid}), 0.6 ({\it dashed})
and 0.8 ({\it dot-dashed}).
These intensity levels roughly correspond to the line core (LC), the mean 
line wing at the mean intensity level between the LC and local continuum, 
and the line wing close to continuum (CLW) respectively.}
\label{plotsix}
\end{figure}

\begin{figure}[!ht]
\includegraphics[width=0.5\textwidth]{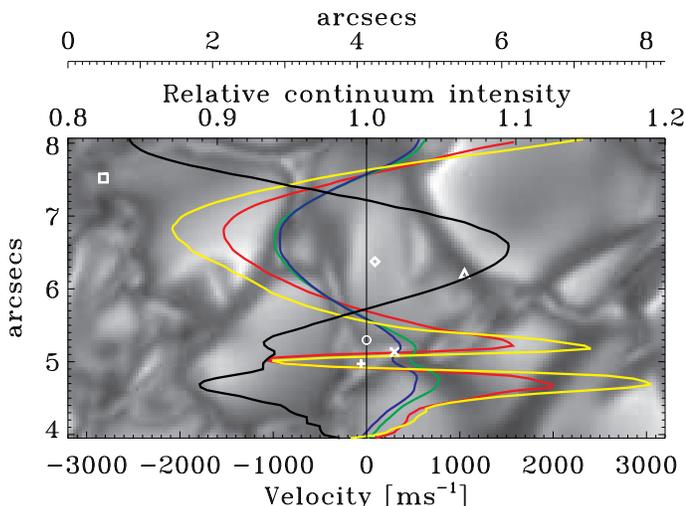}
\caption{
Granule and a ribbon as they appear in the simulations are shown in this 
simulated slit-jaw image. 
The plasma velocity at the height where we have an atmospheric temperature 
equal to the radiation temperature at an intensity level (IL) of 0.65 is 
given in yellow, the non-convolved measured velocity at IL=0.65 
in the Fe~{\small{I}} 5379 {\AA} line is given in
red, while the green and blue lines show the convolved measured velocity at 
IL=0.85 and IL=0.65 respectively in the Fe~{\small{I}} 5379 {\AA}. 
The relative slit continuum intensity is shown in black. 
In Fig.{\ref{plotfive}} we show response functions for different 
solar structures, these structures are defined in this figure: $+$ shows RC, 
$\circ$ shows IGM, $\times$ shows BP, $\Diamond$ shows G, 
$\Box$ shows LMC, and $\triangle$ shows IG.}
\label{plotseven}
\end{figure}

When we measure the velocity in a spectral line at a given intensity level
there are two smearing effects that will influence the measurement. The first effect is an
intrinsic property of the line formation mechanism, namely the finite width of
the response function. The response to a given velocity at a given height in 
a spectral line at a given intensity level is given by the
response function to velocity at this intensity level \citep{1986Magain}. 
We calculate the response functions numerically using a step function in the
velocity, very similar to the way \citet{2005Astrid} calculate their 
response functions to perturbations in temperature.
Typical response functions will have a pronounced peak but 
with a significant width of a few hundred kilometers, 
see Fig.\ref{plotfive}--\ref{plotsix} 
for some examples. In Table~\ref{tyngde} we give average response 
heights (given as the first order moment) in six different solar features
(see Fig.\ref{plotfive} and Fig.\ref{plotseven} for definition).
These response heights are calculated in the line core (LC) and in the
close to continuum line wing (CLW) for the two Fe~{\small{I}} lines
while the moment of the total line (TL) was used for the C~{\small{I}} line.
 
\begin{table}[!ht]
\begin{tabular}{c@{ }c@{ }c@{ }c@{ }c@{ }c}
\multicolumn{6}{c}{First order moment of the response functions.}\\
\tableline
\tableline
&\multicolumn{2}{c}{Fe {\small{I}}~5379 {\AA}}&
\multicolumn{2}{c}{Fe {\small{I}}~5386 {\AA}}&
C {\small{I}}\\%~5380 {\AA}\\
Feature&LC&CLW&LC&CLW&TL\\
&[Km]&[Km]&[Km]&[Km]&[Km]\\
\tableline
RC&232&32&46&-28&2\\
IGM&129&29&78&34&50\\
BP&-36&-112&-44&-116&-134\\
G&305&156&205&141&135\\
LMC&79&-123&-76&-170&-183\\
IG&248&127&211&143&141\\
\tableline
\end{tabular}
\caption{}
\label{tyngde}
\end{table}
 
The width of the response functions means that we do not measure the real 
plasma velocity but rather a height smeared velocity. The difference 
between the response function weighted velocity and the velocity at 
monochromatic optical depth unity can be several hundred m\,s${}^{-1}$, 
see Fig.\ref{plotseven}. 
It is thus important to be aware of this effect when one
talks about Doppler shifts measured in spectral lines. The response function
to velocity in the IGM is quite narrow at all intensity levels,
while the response function in the RC is much wider,
see Fig.\ref{plotsix}. The effect of the response function smearing is,
however, much higher in the IGM since the actual
atmospheric velocity is changing much more quickly with height in IGMs than in 
RCs. This explains why we measure almost the exact plasma velocity in the RC
in Fig.\ref{plotseven} while we measure lower velocities in the IGM.

\begin{figure}[!ht]
\includegraphics[width=0.5\textwidth]{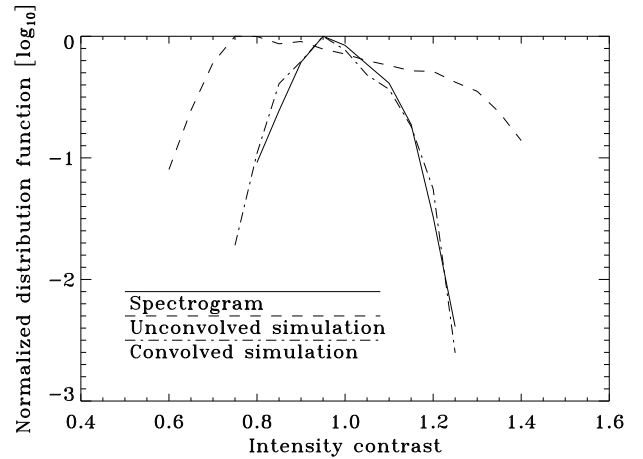}
\caption{To mimic the atmospheric seeing we convolve the simulations with a
linear combination of a Lorentzian and the Airy disk. The resulting NDFs for
the observed spectra ({\it solid}), the unconvolved simulation ({\it dashed}), and the
convolved simulation ({\it dot-dashed}) are shown.}
\label{ploteight}
\end{figure}

\begin{figure}[!ht]
\includegraphics[width=0.5\textwidth]{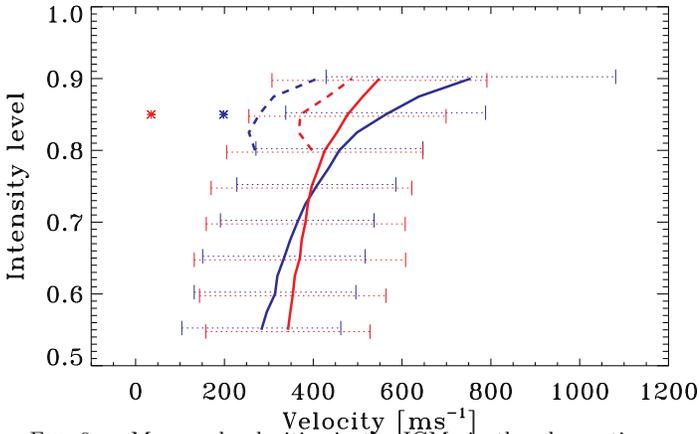}
\caption{Measured velocities in the IGMs in the 
observations ({\it blue}) 
and in the simulations ({\it red}) measured at different intensity levels
for the Fe~{\small{I}} 5379~{\AA} line ({\it solid}),
the Fe~{\small{I}} 5386~{\AA} line ({\it dashed}) and
the C~{\small{I}} line ({\it asterisks}).
The standard deviations for the Fe~\small{I}
\normalsize 5379~{\AA} line are shown with error bars in the respective colors.}
\label{plotnine}
\end{figure}

The second effect is the spatial smearing that originates from
atmospheric seeing, the diffraction pattern of the telescope and
scattering in the atmosphere, the telescope and the spectrograph. Even
if we could determine the total point-spread function (PSF) of the
atmosphere-telescope-spectrograph system, we can't correct for these
effects in the measured spectra since we have spectral information
only along the one dimension of the slit.  Another approach is to
model the PSF and convolve the simulations with such a model to
quantify the effects of the PSF on the measurements.  The PSF of a
partially AO corrected image for short exposures is usually assumed to
consist of a diffraction core and a seeing halo from uncorrected or
partially corrected modes.  We assume that we can use this model for
our spectra even though 80 ms is longer than what is normally accepted
as short exposure.  Furthermore we assume that the seeing halo has the
form of a Lorentzian \citep{1984Nordlund}. 
Our model PSF thus has two free parameters: the width of the Lorenzian
and the fraction of the Lorentzian component compared with the Airy
core. A more detailed model of
the PSF chould not be obtained since we lack the statistics necessary
for the determination of more than two parameters.
The observable is the continuum intensity distribution as
measured in the spectra, excluding the pore. We assume that the
intensity distribution in the simulations is realistic and compare the
normalized distribution function of the intensities in the smeared
simulation with the observed distribution function to determine the
parameters. We find a good fit with a Lorentzian with FWHM of
0\farcs{7} having a fraction of 0.1 at the center of the PSF, see
Fig.\ref{ploteight}. The Strehl ratio of this PSF is 0.15 which is low
compared with what one would expect from the AO system of the SST. One
should remember here that we have no separate component to describe
the scattering of the Earth's atmosphere, which is known to be very
wide, and the Strehl ratio obtained includes this effect. It is also
important to point out that a range of parameters give acceptable fits
to the observed intensity distribution function --- wider Lorentzians
with smaller fraction gives almost the same effect on the normalized
distribution functions as a narrower Lorentzian with larger
fraction. To check the sensitivity of the results to the PSF, we have
performed tests both with a wider central core than the Airy disk and
with different combinations of the width and fraction of the
Lorentzian.  As long as the normalized distribution function is close
to the observed one we do not see significant changes in the resulting
velocities (in comparison with the standard deviations shown in
Fig.\ref{plotnine}).  At the same time, the velocities determined
from the convolved simulations are very different from the velocities
in the unconvolved simulation.  This may seem contradictory but is due
to the fact that intensities and velocities are correlated --- high
intensity granules have large upflows, low intensity intergranular
lanes downflow. When the PSF gives the observed intensity distribution
function this also implies a similar effect on the velocity
distribution function. 
Before comparing with observations, the spectral domain is convolved
with the theoretical resolution of the TRIPPEL spectrograph.

\begin{table}[!ht]
\begin{tabular}{rrrr}
\multicolumn{4}{c}{Central wavelengths}\\
\multicolumn{4}{c}{and convective blueshifts in 
the simulations}\\
\tableline
\tableline
Atomic& Simulation& Reference& Velocity\\
species&[\AA]&[\AA]&[m\,s${}^{-1}$]\\
\tableline
Fe~{\small{I}} & 5379.5728&5379.5740&-69\\ 
C~{\small{I}} & 5380.3143 &5380.3262&-664\\ 
Fe~{\small{I}} & 5386.3301&5386.3341&-223\\
\tableline
\end{tabular}
\caption{}
\label{simwl}
\end{table}

We use the laboratory wavelengths as reference wavelengths for the 
Fe~{\small{I}} lines, while we use the same wavelength as in 
the observations for the C~{\small{I}} line.
Furthermore we calculate the central wavelength in the simulated atlas
and the corresponding blue shift. The results are shown in Table~\ref{simwl}
(with negative velocity meaning blueshift).

Now we can measure the velocities in the IGMs in the simulations. This is done
in a similar manner as we measured the velocities in the IGMs in the 
observations, the results are plotted together with the observed 
results in Fig.\ref{plotnine}. The absolute velocities in the simulations
and the observations differ with 200~m\,s${}^{-1}$ or less, which is amazingly 
close when one consider the uncertainty in the observations of about 
100--200~m\,s${}^{-1}$ and that the standard deviation is about 
200~m\,s${}^{-1}$. Note that we believe that the 
Fe~{\small{I}} 5386~{\AA} line has an error of about 
100~m\,s${}^{-1}$ in the laboratory wavelength. This will increase the 
measured velocities with the same amount.
Furthermore we see an increase in measured velocity with intensity level for 
both the observations and the simulations. This 
increase
is somewhat lower in the simulation than in the observations. 
Note that the C~{\small{I}} line gives substantially lower downflow values than
the iron lines at comparable intensity level. This is true both in the
observations and in the simulations. In the latter, the determined velocity
is completely independent on the reference wavelength used so an error in
the reference wavelength is not the explanation for the low velocities in
the observations. The explanation is rather the differences in the response
functions and the spatial smearing due to the seeing. The difference between
the response functions in the IGM is minimal, but the difference between
the response functions in granules is substantial. In the granule 
the C~{\small{I}} line is formed deeper in the atmosphere, 
and hence this line will show a larger upflow.
When the spatial smearing is applied this difference will lead to a smaller
downflow in IGMs in the C~{\small{I}} line, as observed.
It is clear that the simulations and the observations are not in disagreement,
even though there is room for improvements. 

We also measure the lowest velocities within the magnetic elements
in the simulation. The velocities measured range from about 
200~m\,s${}^{-1}$ upflow to about 200~m\,s${}^{-1}$ down flow. 
The standard deviation is 250--300~m\,s ${}^{-1}$.
Due to the lack of good sampling of bright points in the simulations we
do not include any statistics on bright points in the simulations.

\section{Conclusions}
\label{Con}
We have presented spectrograms of small scale magnetic structures 
with excellent spatial resolution. These spectrograms have been used
to measure the LOS velocities in several different magnetic structures. 
At the edges of the ribbons and flowers we measure downflow with an 
increasing magnitude with depth in the atmosphere. 
The magnitude of the downflow is about 200--700~m\,s${}^{-1}$ with 
increasing velocity deeper down in the atmosphere.
Inside the ribbons and flowers we measure a small upflow.
The upflows are usually in the range 100--300~m\,s${}^{-1}$. 
Our investigation of the velocities inside
bright points shows both up and downflow, 
nevertheless with a mean downflow of a few hundred~m\,s${}^{-1}$. 
In most cases the velocity is closely related to the intergranular 
velocity, which indicate a strong presence of scattered light in the measured 
velocity. 

Furthermore we have investigated the processes that are involved in the 
smearing of the measured velocities. The results from this work can be 
summarized as follows:

1. We have calculated response functions and response heights 
for several different solar features. The width of the response functions
is usually a few hundred km.

2. We have convolved the simulations with a simple model of the point spread
function of the observations. Due to the strong horizontal gradients in
the velocity field in and close to small scale magnetic features, this
spatial smearing has a strong effect on the measurements:
the measured velocities in the intergranular lanes
close to magnetic elements are decreased with about 1~km\,s${}^{-1}$.

3. Since the simulations are reproducing the observations reasonably well
we believe that the plasma velocities in the simulations are close to the
real solar plasma velocities.
In particular the velocities in and close to larger magnetic structures such
as ribbons are well reproduced.
The velocity in the simulation in intergranular lanes close to magnetic 
structures increases from about 1.5~km\,s${}^{-1}$ at 150~km height to
3.3~km\,s${}^{-1}$ at 0~km.
 
The velocity inside bigger magnetic structures such as ribbons in the 
simulations is about 1--2~km\,s${}^{-1}$ upflow, with the velocity 
changing only a few hundred~m\,s${}^{-1}$  
throughout the formation height between about $-$100~km and 250~km. 
\acknowledgements
\emph{Acknowledgments.}
This research was supported by the European Community's Human Potential
Program through the ESMN (contract HPRN-CT-2002-00313)
and TOSTISP (contract HPRN-CT-2002-00310) programs and by The 
Research Council of Norway through grant 146467/420 and through
grants of computing time from the Programme for Supercomputing.
The simulation calculations were supported
by NASA grants NNG04GB92G and NAG 5-12450, and NSF grant AST 0205500.  
The calculations were performed at NASA's High End Computing
Program Columbia computer system, at the National Center for Supercomputer 
Applications, which is supported by the National Science Foundation, 
and at Michigan State University.
The Swedish 1-m Solar Telescope is operated on the island of La Palma
by the Institute for Solar Physics of the Royal Swedish Academy of
Sciences in the Spanish Observatorio del Roque de los Muchachos of the
Instituto de Astrof{\'\i}sica de Canarias.
This research has made use of NASA's Astrophysics Data System.

\bibliographystyle{apj}
\bibliography{ms}

\end{document}